\documentclass[12pt,journal,onecolumn]{IEEEtran}
\RequirePackage[normalem]{ulem} 
\RequirePackage{color}\definecolor{RED}{rgb}{1,0,0}\definecolor{BLUE}{rgb}{0,0,1} 

\RequirePackage{color}

\usepackage{tikz}
\usepackage[sumlimits,intlimits]{amsmath}
\usepackage{amsmath,amsfonts,amssymb}%
\usepackage{bm}%
\usepackage{graphicx,graphics}%
\usepackage{cases}%
\usepackage[noadjust]{cite}%
\usepackage{color}%
\usepackage{cite,url}
\usepackage{verbatim}
\usepackage{enumerate}
\usepackage{algorithm}
\usepackage{algorithmic}
\usepackage{balance}
\usepackage{multirow}
\usepackage{stfloats}
\usepackage{subfigure}
\usepackage{xspace}
\usepackage{smartref}
\usepackage{epstopdf}
\usepackage{setspace}

\RequirePackage[normalem]{ulem} 
\RequirePackage{color}\definecolor{RED}{rgb}{1,0,0}\definecolor{BLUE}{rgb}{0,0,1} 

\begin{document}

\title{\vspace*{-14pt}
\huge{Interference Mitigating Satellite Broadcast Receiver using Reduced Complexity List-Based Detection in Correlated Noise}
\\
}
\author{
{Zohair Abu-Shaban, ~\IEEEmembership{Student Member, IEEE}, Hani Mehrpouyan,~\IEEEmembership{Member, IEEE},\\ Bhavani Shankar M. R, ~\IEEEmembership{Member, IEEE}, and Bj\"{o}rn Ottersten, ~\IEEEmembership{Fellow, IEEE.}
}
}
\maketitle
\thispagestyle{empty}
{\let\thefootnote\relax\footnotetext{
Zohair Abu-Shaban,  Bhavani Shankar R and Bj\"{o}rn Ottersten are with the Interdisciplinary Centre for Security, Reliability and Trust at the University of Luxembourg, Luxembourg.  Hani Mehrpouyan is with the Department of Electrical Engineering at the University of California, Riverside. Emails: zohair.abushaban@uni.lu, hani.mehr@ieee.org, bhavani.shankar@uni.lu, bjorn.ottersten@uni.lu. \\ This work is supported by the National Research Fund (FNR), Luxembourg. Project ID: 4043055.\\ Part of this work is accepted for publication in the proceedings of the IEEE International Conference on Communications (ICC), June 2014, Sydney Australia.
} 
\vspace{-1cm}
\doublespacing
\begin{abstract}
The recent commercial trends towards using smaller dish antennas for satellite receivers, and the growing density of broadcasting satellites, necessitate the application of robust adjacent satellite interference (ASI) cancellation schemes. This orbital density growth along with the wider beamwidth of a smaller dish have imposed an \textit{overloaded} scenario at the satellite receiver, where the number of transmitting satellites exceeds the number of receiving elements at the dish antenna. To ensure successful operation in this practical scenario, we propose a satellite receiver that enhances signal detection from the desired satellite by mitigating the interference from neighboring satellites. Towards this objective, we propose a reduced complexity list-based group-wise search detection (RC-LGSD) receiver under the assumption of spatially correlated additive noise. To further enhance detection performance, the proposed satellite receiver utilizes a newly designed whitening filter to remove the spatial correlation amongst the noise parameters, while also applying a preprocessor that maximizes the signal-to-interference-plus-noise ratio (SINR). Extensive simulations under practical scenarios show that the proposed receiver enhances the performance of satellite broadcast systems in the presence of ASI compared to existing methods.
\end{abstract}
\vspace{-3mm}


\vspace{-1cm}
\section{Introduction}
\subsection{Motivation}
Over the past decade, satellite broadcast services including, direct-to-home (DTH), have shown significant growth and are expected to continue to represent a principal sector of the overall satellite business in the future \cite{SatIndustry}. To meet the needs of satellite broadcast market, more satellites are launched and typically stationed in the geostationary orbit (GEO). As a result of this higher satellite density and use of common frequency bands amongst these satellites, e.g., the Ku-band, the receivers are more susceptible to \textit{adjacent satellite interference} (ASI) \cite{Elbert2004}. In addition, it is commercially attractive for home users to utilize satellite receivers with small-aperture antennas due to their reduced manufacturing and mounting costs. However, it is well known that a smaller dish size has a wider radiation pattern resulting in reduced directivity and higher levels of ASI at the receiver. These two factors have created the need for designing new algorithms that can more effectively mitigate the ASI. Such algorithms are expected to enhance the throughput of satellite receivers and provide the satellite broadcasting industry with an edge over other existing alternatives, e.g., cable and fiber optics.

A fixed satellite receiver can benefit from the known location of the satellites by employing a multiple-feed antenna, known as \textit{multiple low noise blocks} (MLNBs). The number of LNBs is usually limited to 2$-$3 feeds \cite{grotz2010} to reduce hardware costs, mechanical support requirements, and electromagnetic blockage. The motivation to consider an overloaded system, i.e., a system with a higher number of satellites than MLNBs, stems from the limited number of LNBs compared to the larger number of the satellites that fall within the view of the wider radiation pattern of a smaller dish antenna. Fig.~\ref{fig:setup_figure} illustrates a conceptual setup of this practical scenario.

\begin{figure}[!t]
\centering
\begin{tikzpicture}
\tikzstyle{every node}=[font=\tiny]
\node[above right] (img) at (0,0) {\includegraphics[scale=.75]{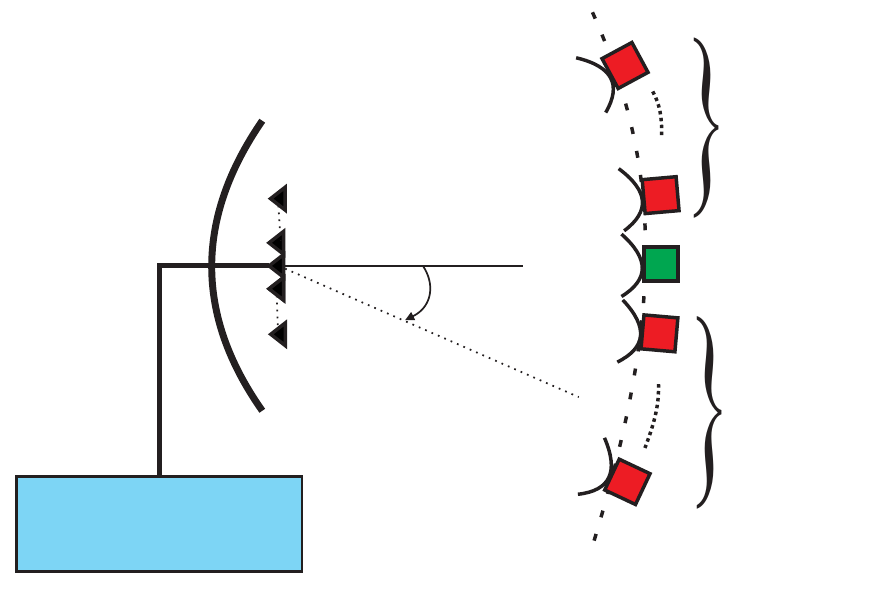}};
\node at (40pt,20pt) {\normalsize $Receiver$};
\node at (90pt,69pt) {\footnotesize $\theta_n$};
\node at (76pt,100pt) {\footnotesize $M$ feeds};
\node at (190pt,107pt) {\footnotesize interference};
\node at (190pt,43pt) {\footnotesize interference};
\node at (190pt,75pt) {\footnotesize desired satellites};
\node at (135pt,75pt) {\footnotesize $s_1$};
\node at (134pt,60pt) {\footnotesize $s_2$};
\node at (134pt,88pt) {\footnotesize $s_3$};
\node at (129pt,110pt) {\footnotesize $s_N$};
\node at (122pt,29pt) {\footnotesize $s_{N-1}$};
\node at (122pt,139pt) {\footnotesize $N$ satellite};
\node at (125pt,10pt) {\footnotesize GEO};
\end{tikzpicture}
\vspace{-4mm}
        \caption{The system setup for $N$ satellites in the geostationary orbit, and $M$ LNBs. The dish is directed to the desired satellite, $s_1$.}
        \label{fig:setup_figure}
        \vspace{-4mm}
\end{figure}
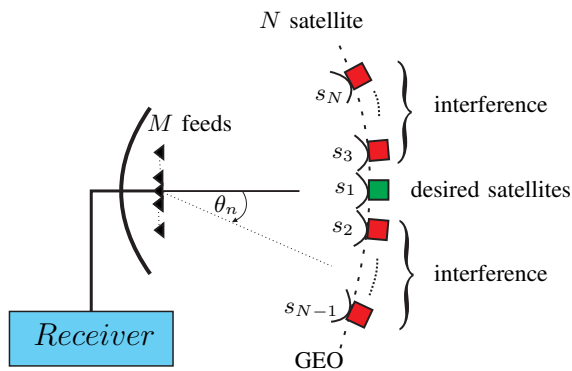

\subsection{Prior Work}
Interference cancellation for signals arising from different satellites and having partial frequency overlap is addressed in \cite{beidas2002}. Subsequently, this work is extended in \cite{Schwarz2007} to support television signals conforming to the digital video broadcasting standard, DVB-S2 \cite{dvbb-s2}. However, these two works employ a single input receiver. Consequently, they do not exploit spatial processing for interference cancellation. On the other hand, an MLNB-based two-stage satellite receiver is proposed in \cite{grotz2010}. This receiver applies a linear preprocessing stage that minimizes the overloading effect on the detection process, followed by a non-linear iterative detection stage. However, the approach in \cite{grotz2010} exhibits a poor bit error rate performance for quadrature phase-shift-keying (QPSK) signals. Hence, the performance of the receiver in \cite{grotz2010} is expected to be further degraded for the higher order modulations that are under consideration in this work.

ASI mitigation in \textit{full} frequency overlapping, i.e., co-channel, satellite broadcast systems is addressed in \cite{AbuShaban2012}. While considering the satellite position relative to the antenna orientation, this approach applies a successive interference canceller (SIC) along with two different beamforming methods to mitigate the impact of ASI. The aim of \cite{AbuShaban2012} is to detect the signals from as many satellites as the number of the LNBs that are available at the receiver. Although the algorithm shows an acceptable bit-error rate performance for QPSK satellite signals, the performance greatly deteriorates as the modulation order increases.

For non-satellite scenarios, multi-user detection (MUD) and interference cancellation techniques for \textit{overloaded} receivers are discussed in \cite{Bayram2000,krause2011,hicks2001,kapur2003,Colman2008,Grant1998}. The approaches of these works are generally based on either the maximum-likelihood approach or its lower complexity variations. In \cite{Bayram2000}, while considering co-channel signals in an overloaded scenario, it is shown that the joint maximum-likelihood (JML) detector is an optimal detector. The drawback of the JML detector is that its complexity grows exponentially with the number of received signals and the modulation order. This implies that for modulations schemes, such as 8 phase-shift keying (8PSK) and 16 amplitude and phase-shift keying (16APSK) that are used in DVB-S2, the receiver becomes unfeasible for the end-user equipment.

\subsection{Conventional List-Based Receiver}
A two-stage receiver that employs a reduced complexity search algorithm known as \textit{list-based group search detection} (LGSD) is proposed in \cite{krause2011}. LGSD is designed to search over a smaller space compared to the JML detector by estimating a list of highly probable candidates. The first stage of the receiver is a \textit{linear preprocessor} that preconditions the received signal by maximizing the output signal-to-noise ratio (SNR) using maximum ratio combining (MRC). The second stage of the receiver is a non-linear detector that is composed of two processes. The first process creates a relatively short candidate list that is used by the second process to carry out JML detection. In essence, LGSD partitions the channel matrix into lower dimensional search spaces, and the received vector into sub-vectors before executing JML detection on these sub-vectors. Iterating between the two processes improves the performance with some added complexity. Although LGSD can reduce the complexity of the detection process in overloaded scenarios, the approach in \cite{krause2011} is not directly applicable to satellite broadcasting scenarios due to the limitations outlined below.

In \cite{krause2011}, the interference at the receiver is modelled as a white Gaussian process for the diversity combining, while the additive noise is assumed to be uncorrelated. These two assumptions may not hold for the signal detection in satellite systems \cite{grotz2010,AbuShaban2012,AbuShaban2013}. Thus, we build our work on the conventional LGSD receiver and address its shortcomings by focusing on modifying these two assumptions to suit satellite broadcasting systems. Furthermore, we introduce a truncation procedure in the second stage to further reduce the computational complexity of the LGSD approach.

\subsection{Contributions}
In this paper, we design an overloaded multiple input receiver for broadcast satellite systems. The receiver is assumed to use a small-size antenna, e.g., $<$40 cm, that is equipped with multiple LNBs. As shown in Fig.~\ref{fig:setup_figure}, the dish is assumed to be fixed and directed towards the central satellite, which we refer to as the \textit{desired} satellite. The remaining satellites in the view of the antenna are also assumed to be operating in the same frequency band and are referred to as the \textit{interfering} satellites. Due to the small dish size, the antenna patterns are wide, causing a high level of interference. The contributions of this paper can be summarized as:

\begin{itemize}
  \item The knowledge of the satellite location and the fixed antenna setup is used to accurately model the interference from neighboring satellites instead of treating it as additive white noise. Subsequently, an efficient beamformer based on the signal-to-interference-plus-noise (SINR) maximization criterion is utilized.
  \item We proposed to use a practical model of the additive noise that takes into account the correlation amongst the LNBs. Due to the overlapping patterns of the MLNBs, one LNB pattern affects the neighboring LNBs. Hence, the additive noise at the LNBs are spatially correlated \cite{grotz2010}. The characteristics of this spatial correlation is obtained from the radiation patterns of the LNBs. Subsequently, a new whitening filter is derived that is better suited to the proposed beamformer and accurately models the correlated noise in satellite systems.

  \item Due to the rank deficiency of the overloaded satellite system under consideration here, the equivalent channel matrix as seen by the detector, consists of a number of rows whose entries are zeros. Thus, we have proposed to truncate the channel matrix by removing these rows to reduce the overall complexity of a satellite receiver. This complexity reduction stems from the reduction in the number of calculations that are required to create the candidate list for the LGSD algorithm.

   \item Using the proposed beamforming scheme, noise whitening filter, and channel matrix truncation procedure, a new receiver structure denoted by \emph{Reduced Complexity-LGSD} (RC-LGSD) is proposed that can be applied to overloaded satellite broadcasting systems. Next, extensive Monte-Carlo simulations are carried out to investigate the performance of the proposed receiver in realistic satellite broadcast scenarios. These simulations show that the proposed receiver structure can outperform the algorithm in \cite{krause2011} for satellite broadcasting systems. We also show that the proposed receiver is less complex than our previous work presented in \cite{AbuShaban2013}.

 \item An in-depth investigation of the trade-off between performance and complexity for the proposed receiver is carried out. Moreover, the receiver performance in the presence of pointing errors is also evaluated.
\end{itemize}
\subsection{Notations}
Italic lowercase and uppercase letters, e.g., \textit{a} and \textit{A} are used to denote scalars, while  bold lowercase and uppercase letters, e.g., \textbf{a} and \textbf{A}, are used to denote column vectors and matrices, respectively. ${a}_m$ and $\textbf{A}(m)$ refer to the $m^{th}$ element of vector $\textbf{a}$ and the $m^{th}$ row of a matrix $\textbf{A}$, respectively. $\textbf{a}_n$ represents the $n^{th}$ column of a matrix $\textbf{A}$. The transpose and Hermitian transpose are denoted by  $(\cdot)^T$ and $(\cdot)^H$, respectively. The pseudo-inverse of matrix $\mathbf{A}$ is denoted by $\mathbf{A}^\dag$. $a^{*}$ denotes the complex conjugate of $a$. $\mathcal{C}^N$ denotes the $N$-dimension complex space and $\textbf{I}_N$ is used to indicate the $N\times{N}$ identity matrix. $\phi$ denotes the empty set, while $\mid{\Gamma}\mid$ represents the cardinality of a set $\Gamma$. The Euclidean distance of the vector $\textbf{a}$ and the Frobenius norm of matrix $\textbf{A}$ are denoted by $\|\textbf{a}\|^2$ and $\|\textbf{A}\|_F$, respectively. Finally, $\mathop{\mathbb{E}}\left[\cdot\right]$ denotes the expectation operator.

\subsection{Outline}
Section \ref{sec:section-sys-model} highlights the system model, the considered scenario, and the underlying assumptions. Section \ref{sec:section-prop_rcv} describes the proposed RC-LGSD detector and presents the proposed preprocessor including the beamformer and the noise whitening filter. The complexities of the proposed and existing algorithms are analyzed in Section \ref{sec:complexity}. The simulation environment and results are discussed in Section \ref{sec:sim}, while Section \ref{sec:conc} concludes the paper.

\section{Assumptions and Signal Model }\label{sec:section-sys-model}
Let us consider $N$ adjacent satellites orbiting the GEO and broadcasting to an overloaded receiver equipped with a small-size dish and $M$ LNBs. The following assumptions are made throughout this paper:
\begin{itemize}
\item\emph{Overloaded receiver ($N>M$)}: Due to practical factors such as cost reduction and electromagnetic blockage prevention, the number of LNBs, $M$, is to be kept small, i.e., 2$-$3 LNBs. For small-aperture reflectors, a larger number of satellites fall within the field-of-view of the antenna. Depending on the dish diameter, $D$, and the operating wavelength $\lambda$, the reflector $3$-dB beamwidth can be estimated by $(70\lambda/D)^\circ$ \cite{SatPrinciples}. The number of satellites can then be estimated knowing that the GEO satellites are usually separated by an angular spacing of $2.5^\circ-3^\circ$ \cite{grotz2010}. For example, the $3$-dB beamwidth of the central LNB of a dish with a diameter of $35$ cm operating in the Ku-band is $5^\circ-6^\circ$. Thus, one can expect $3$ satellites to fall within the field-of-view of a single LNB dish. Adding more LNBs extends the field-of-view and more satellites can be observed at the receiver. However, adding more LNBs offers more degrees of freedom, which is beneficial in the joint detection process. Fig.~\ref{fig:pattern} shows an example of the antenna patterns for 35-cm dish equipped with 3 LNBs. It can be seen that with the central LNB, 3 satellites are observed at the receiver. Moreover, although 5 satellites are observed at the receiver when using two more LNBs, the provided receive diversity enables us to apply a beamforming scheme that improves the receiver performance.

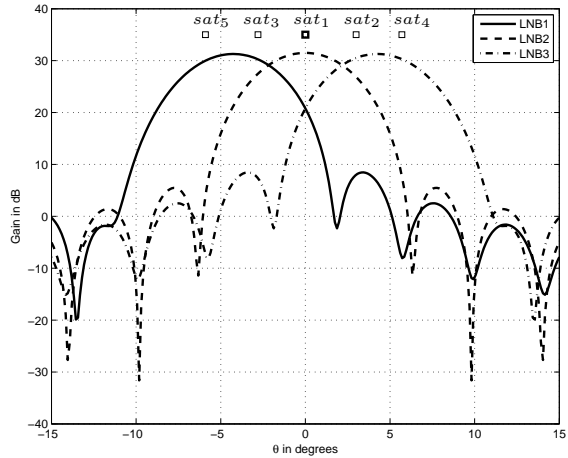
\begin{figure}[!t]
\centering
\begin{tikzpicture}
\tikzstyle{every node}=[font=\tiny]
\node[above right] (img) at (0,0) {\includegraphics[scale=0.45, trim=0cm 0cm 0cm 10mm, clip=true]{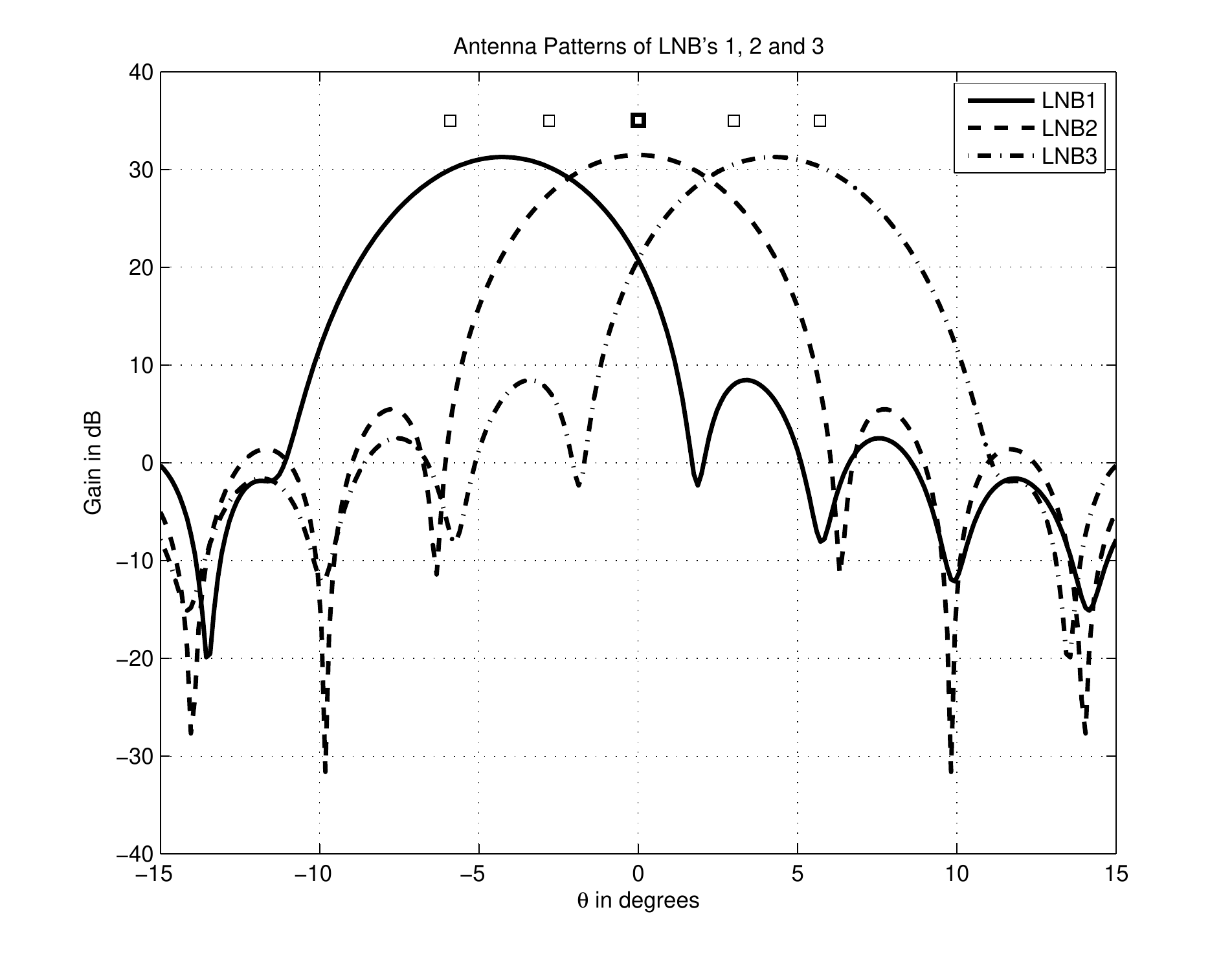}};
\node at (135pt,178pt) {$sat_1$};
\node at (116pt,178pt) {$sat_3$};
\node at (97pt,178pt) {$sat_5$};
\node at (154pt,178pt) {$sat_2$};
\node at (173pt,178pt) {$sat_4$};
\end{tikzpicture}
\vspace{-7mm}
        \caption{Radiation patterns of three LNBs mounted on a 35-cm dish.}
        \label{fig:pattern}
\end{figure}

   \item \emph{The system is assumed to be synchronized:} Although the LNBs can use the same oscillator to reduce the frequency and phase uncertainties, the received signals are assumed to be symbol-synchronized. The synchronization parameters are assumed to be supplied by a synchronizer block at the digital front-end of the receiver. Such an assumption has been also made in prior work in \cite{AbuShaban2012,AbuShaban2013,Grotz2005}.
  \item \emph{Spatially correlated additive noise:} Similar to prior work in \cite{AbuShaban2012,AbuShaban2013,Grotz2005}, the radiation patterns of the MLNBs are assumed to be partially overlap. This pattern overlap induces spatial correlation in the noise emanating from other sources such as the gateway and satellite components.

  \item \emph{The signals are assumed to comply with the DVB-S2 standard and are independently transmitted:} Signal parameters such as modulation and power level can be estimated by the receiver using the frame structure defined in DVB-S2. Since these parameters are already needed for synchronization, we assume that they are provided by the synchronization block. Such an assumption has been also made in prior work in \cite{Grotz2005}.
   \item \emph{The channel is assumed to be known and fixed:} A line-of-sight link and a clear sky are assumed. Therefore, the channel is mainly dependent on parameters such as the antenna geometry and electrical specifications such as diameter, focal length, oscillator stability, low noise amplifier gain, etc. Since these parameters do not vary quickly, they are assumed fixed over the transmission interval. Accordingly, the channel is expressed as a function of the satellite location angles, which can be estimated by the knowledge of the antenna radiation patterns. Initially, ideal channel is assumed, before considering the pointing error, where performance is evaluated in the presence of pointing angle uncertainty.
\end{itemize}
Under the above assumptions, the baseband symbol-sampled received signal vector at the output of the synchronizer is given by
\begin{align}\label{eq:sys-eq}
\mathbf{r}[k] = \mathbf{A} \mathbf{s}[k]+\mathbf{n}[k],
\end{align}
where
\begin{itemize}
\item $\mathbf{r}[k]\triangleq\big[{r}_1 [k], {r}_2 [k], ..., {r}_M [k]\big]^T$ is the received symbol vector at time instant $k$,
\item $\mathbf{A}\triangleq[A_{i,j}]$ is an $M\times{N}$ matrix representing the antenna array response with $A_{i,j}$ denoting the complex gain of the $i^{th}$ LNB in the direction of the $j^{th}$ satellite,
\item $\mathbf{s}[k]\triangleq\big[s_1 [k], s_2 [k], ..., s_{N}[k]\big]^T$ is the transmitted symbol vector, where $s_j [k]$ is drawn from a zero-mean unit-variance signal constellation, $\omega$ of cardinality $K$, and $s_1[k]$ corresponds to the desired satellite shown in Fig.~\ref{fig:setup_figure}, and
\item $\mathbf{n}[k]\triangleq\big[n_1[k], n_2 [k], ..., n_M [k]\big]^T$ is the additive noise vector that is modelled as a zero-mean Gaussian process with covariance matrix $\mathbf{R}_{nn}=\sigma_n^2\mathbf{K}_{nn}$.
\end{itemize}

Here, $\sigma_n^2$ is the noise variance and $\mathbf{K}_{nn}\triangleq{K_{i,j}}$ is the matrix of normalized correlation coefficients. Note that $\mathbf{K}_{nn}$ is a function of the radiation patterns of the LNBs and is determined by the magnitude of overlapping amongst these patterns. To obtain $K_{i,j}$, denote the square root of the radiation pattern of the $m^{th}$ LNB by $p_m(\theta)$, then
\begin{align}\label{eq:matrix-K}
K_{i,j}=\frac{\int_{-\pi}^{\pi}p_i(\theta)p_j^{*}(\theta) d\theta}{\sqrt{\int_{-\pi}^{\pi}p_i(\theta)p_i^{*}(\theta)d\theta}{\sqrt{\int_{-\pi}^{\pi}p_j(\theta)p_j^{*}(\theta)d\theta}}},
\end{align}
where $\theta$ is in radians \cite{Hon2009}. Eq. (\ref{eq:matrix-K}) illustrates the relationship between the overlapping LNB patterns and the spatial correlation amongst the additive noise at each LNB. In other words, the wider the overlapping, the higher the spatial correlation amongst the additive noise parameters corresponding to each LNB. This implies that for a small dish with a wide pattern, the additive noise parameters at the outputs of the LNBs have stronger correlations.

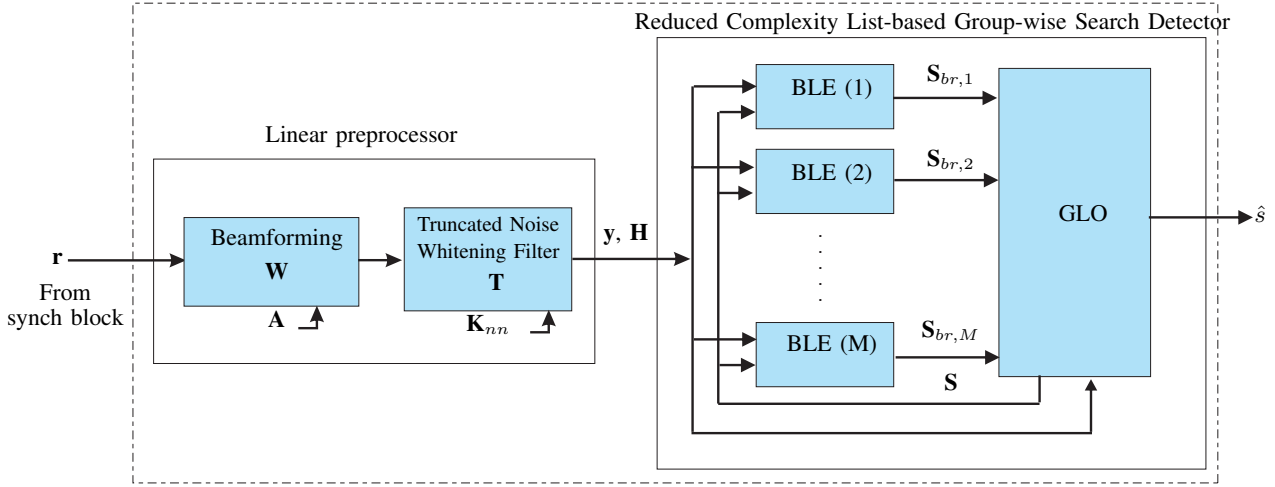
\begin{figure*}[!t]
\centering
\begin{tikzpicture}
\tikzstyle{every node}=[font=\footnotesize]
\node[above right] (img) at (0,0) {\includegraphics[scale=.45]{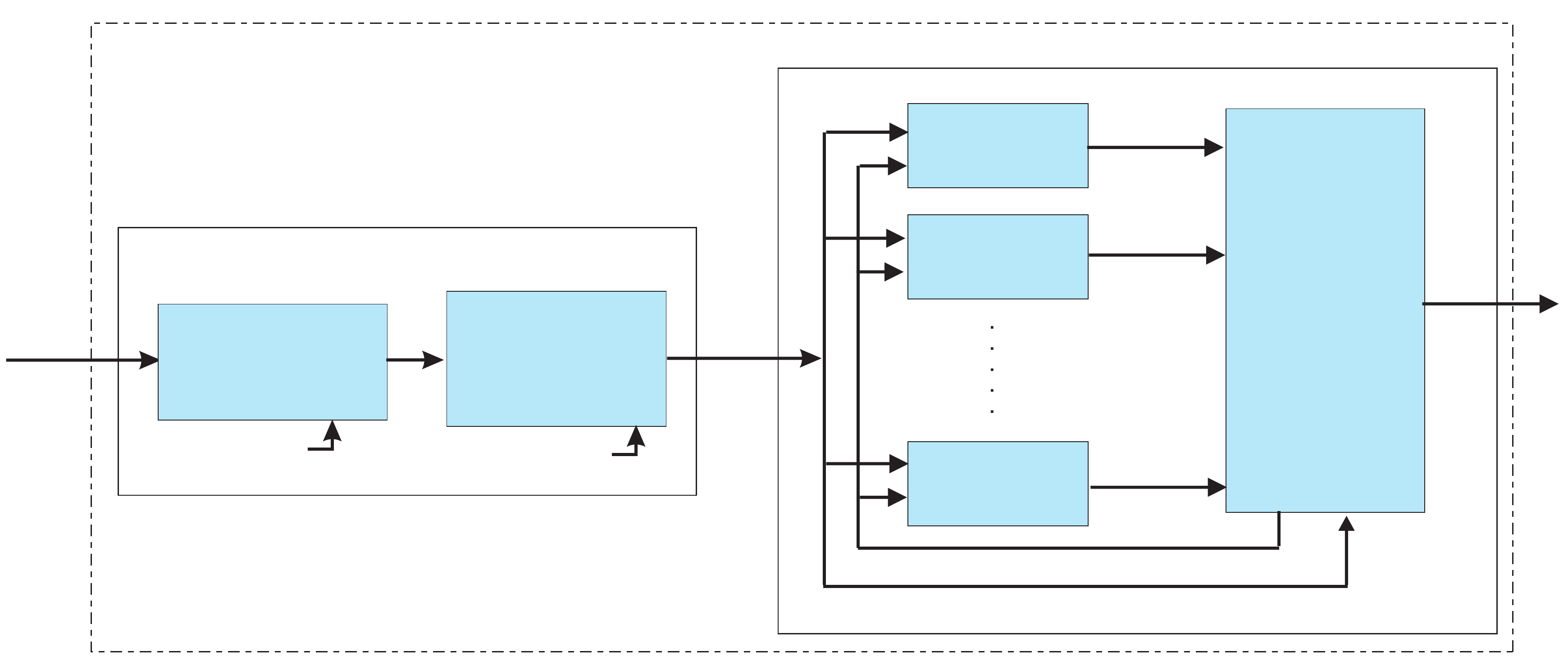}};
\node at (117pt,139pt) {Linear preprocessor};
\node at (85pt,70pt) {$\textbf{A}$};
\node at (85pt,88pt) {$\textbf{W}$};
\node at (85pt,100pt) {Beamforming};
\node at (165pt,106pt) {\scriptsize Truncated Noise};
\node at (165pt,95pt) {\scriptsize Whitening Filter};
\node at (165pt,68pt) {$\textbf{K}_{nn}$};
\node at (168pt,84pt) {$\textbf{T}$};
\node at (2pt,93pt) {\textbf{r}};
\node at (5pt,80pt) {From};
\node at (5pt,70pt) {synch block};
\node at (217pt,102pt) {\textbf{y}, \textbf{H}};
\node at (333pt,182pt) {Reduced Complexity List-based Group-wise Search Detector};
\node at (295pt,157pt) {BLE (1)};
\node at (295pt,125pt) {BLE (2)};
\node at (295pt,60pt) {BLE (M)};
\node at (340pt,162pt) {$\textbf{S}_{br,1}$};
\node at (340pt,130pt) {$\textbf{S}_{br,2}$};
\node at (340pt,65pt) {$\textbf{S}_{br,M}$};
\node at (340pt,46pt) {\textbf{S}};
\node at (390pt,110pt) {GLO};
\node at (457pt,110pt) {$\hat{s}$};
\end{tikzpicture}
\vspace{-4mm}
        \caption{The block diagram of the proposed receiver. BLE: Branch list estimator. GLO: Global list optimizer.}
        \label{fig:blockDiagram}
\end{figure*}

\section{Proposed Overloaded Receiver Design}\label{sec:section-prop_rcv}
In this section, we discuss the design of the proposed overloaded satellite receiver. The block diagram of this two-stage receiver is illustrated in Fig.~\ref{fig:blockDiagram}. As shown in Fig.~\ref{fig:blockDiagram}, the receiver consists of two stages:
\begin{enumerate}
\item The linear preprocessor stage, which is composed of a beamformer followed by a noise whitening filter. The objective of this stage is to reduce the overloading effect on the receiver performance by better optimizing the output with respect to the SINR criterion, while alleviating the impact of the correlated noise on the detection process.

\item The second stage of the receiver is the so-called LGSD algorithm, which itself comprises two steps:
\begin{enumerate}
\item In the first step of the LGSD algorithm, a bank of branch list estimators (BLE) simultaneously nominate multiple candidate lists that are used in the second stage of the algorithm.
\item In the second stage, the global list optimizer (GLO) uses the candidate lists supplied by the BLEs to determine the best ML estimate of the transmitted symbol vector. Finally, the output list of GLO is fed back to the BLEs for further iterations.
\end{enumerate}
\end{enumerate}

\subsection{Linear Preprocessor}
Let us denote the output of the linear preprocessor by $\textbf{r}'$. By the dropping the time index $k$, $\textbf{r}'$ can be written as
\begin{align}
\mathbf{r}' = \mathbf{F}^H\mathbf{W}^H\mathbf{r}\triangleq\mathbf{A}' \mathbf{s}+\mathbf{n}',
\end{align}
where $\textbf{W}$ and $\textbf{F}$ are the beamforming and noise whitening filter matrices of size $(M\times{N})$ and $(N\times{N})$, respectively. $\textbf{A}'\triangleq\textbf{F}^H\textbf{W}^H\textbf{A}$ represents the equivalent channel matrix at the input of the detector and $\mathbf{n}'=\mathbf{F}^H\mathbf{W}^H \mathbf{n}$ is the whitened additive Gaussian noise process at the output of the preprocessor. In the following subsections we present the process for designing the matrices \textbf{W} and \textbf{F}.
\subsubsection{Beamforming}
Although the original LGSD algorithm proposed in \cite{krause2011} uses an SNR-based beamformer, i.e., an MRC beamformer, in satellite systems the structure of the interference can be used to design an SINR-based beamformer that can further enhance the performance of the receiver compared to an MRC approach. This is even more important since future satellite systems with small-size receiving dishes maybe more interference-limited than noise-limited. This conclusion is also supported by the simulation results presented in Section \ref{sec:sim}. From \cite{krause2011}, the MRC beamformer is defined as
\begin{align}
\mathbf{W} = \mathbf{A}.
\end{align}
Writing \textbf{A} and \textbf{r} in terms of the columns of \textbf{A} leads to
\begin{align}
\mathbf{A} \triangleq \left[\mathbf{a}_1, \mathbf{a}_2, ... \mathbf{a}_N \right],
\end{align}
\begin{align}
 \mathbf{r}=\sum_{m=1}^{N}\mathbf{a}_{m}s_{m}+\mathbf{n}.
\end{align}

Consequently, the beamforming vector that maximizes the SINR for the $m^{th}$ stream, is given by
\begin{align}\label{eq:MRC_BF}
  \mathbf{w}_m \triangleq arg \max_{\mathbf{w}\in{\mathcal{C}^M}}\frac{\mathbf{w}^H\mathbf{R}_m\mathbf{w}}{\mathbf{w}^H(\mathbf{R}-\mathbf{R}_m)\mathbf{w}}, \hspace{8pt}1\leq{m}\leq{N},
\end{align}
where $\mathbf{R}=\mathop{\mathbb{E}}\left[ \mathbf{r} \mathbf{r}^H\right]$ and $\mathbf{R}_m=\mathbf{a}_{m}\mathbf{a}_{m}^H$. The solution of this optimization problem, known as the\textit{ generalized Rayleigh quotient}, is obtained by solving the generalized eigenvalue problem. Hence, $\mathbf{w}_m$ is the eigenvector corresponding to the maximum eigenvalue of $(\mathbf{R}-\mathbf{R}_m)^{-1}\mathbf{R}_m$ \cite{eigen}. Alternatively, it can be shown that $\mathbf{w}_m$ can be calculated directly as a Wiener-Hopf beamformer given by \cite{VanTrees}
\begin{align}\label{eq:MRC_BF}
\mathbf{w}_m = \mathbf{R}^{-1}\mathbf{a}_m.
\end{align}
Accordingly,
\begin{align}
\mathbf{W} \triangleq \left[\mathbf{w}_1, \mathbf{w}_2, ... \mathbf{w}_N \right]=\mathbf{R}^{-1}\mathbf{A}.
\end{align}
\subsubsection{Noise Whitening Filter}\label{sec:truncation_noise}
Recall that the additive noise at the receiver is spatially correlated due to the overlapping patterns of the LNBs. Thus, in this section we design a whitening filter for satellite systems that whitens the noise by taking into account the spatial correlation amongst the LNBs.

The covariance matrix of the noise vector $\textbf{n}'=\mathbf{F}^H\mathbf{W}^H\mathbf{n}$ is given by
\begin{align}\label{eq:noise_corr}
\mathbf{R}_{n'n'}= \mathop{\mathbb{E}}\left[ \mathbf{n}' \mathbf{n}'^H\right] = \sigma^2_n\mathbf{F}^H\mathbf{W}^H \mathbf{K}_{nn}\mathbf{WF}.
\end{align}
Note that for the overloaded scenario under consideration, $\mathbf{R}_{n'n'}$ is rank deficient. Our goal is to find the noise whitening filter matrix, \textbf{F} such that it minimizes $\|\mathbf{F}^H\mathbf{GF}-\mathbf{I}_N\|_F$, where $\mathbf{G}=\mathbf{W}^H \mathbf{K}_{nn}\mathbf{W}$. Since $\mathbf{K}_{nn}$ is a covariance matrix, $\mathbf{G}$ is a positive semi-definite matrix. Hence, it can be written via the eigenvalue decomposition as $\mathbf{G}=\mathbf{UL}\mathbf{U}^H$. Consequently, it is straightforward to show that a solution of $\mathbf{F}$ can be determined as
\begin{align}
\mathbf{F}=\mathbf{U}\left(\mathbf{L}^\dagger\right)^{\frac{1}{2}}.
\end{align}
Since $\mathbf{W}$ has a size of $(M\times{N})$, the matrix $\mathbf{G}$ is rank deficient. Consequently, the rank of $\mathbf{L}$ is $M$. In other words, the last $(M-N)$ columns of $\mathbf{F}$ are zeros. Hence, the output of the noise whitening filter is independent of the last $(M-N)$  elements in the input vector. Consequently, these columns can be removed, resulting in a truncated noise whitening filter $\mathbf{T}$ of size $(N\times{M})$. This implies that, after the truncation operation, the equivalent channel matrix at the input of the detector and the whitened noise vector are given by
\begin{align*}
\mathbf{H}&\triangleq\mathbf{T}^H\mathbf{W}^H\mathbf{A},\\
\mathbf{z}&\triangleq\mathbf{T}^H\mathbf{W}^H\mathbf{n},
\end{align*}
respectively.
Note that $\textbf{R}_{zz}= \mathop{\mathbb{E}}\left[ \textbf{z} \textbf{z}^H\right]=\sigma_n^2\textbf{I}_M$, which verifies that the output of the preprocessor is uncorrelated.

Based on the application of beamforming and noise whitening processes, the input-output relation at the output of the preprocessor can be written as
\begin{align}
\mathbf{y}=\mathbf{H}\mathbf{s}+\mathbf{z}.
\end{align}
\subsection{Multi-User Detection}
\subsubsection{Joint Maximum-Likelihood (JML) Detection}
We now consider the JML detector of \textbf{s}. Starting from the likelihood function given by
\begin{align}
f\left(\mathbf{y}\mid\textbf{H},\textbf{s}\right)=\frac{1}{\left(2\pi\sigma_n^2\right)^M}\exp\left(-\frac{\|\mathbf{y}-\mathbf{H}\mathbf{s}\|^2}{\sigma_n^2}\right),
\end{align}
and noting that maximizing $f\left(\mathbf{y}\mid\textbf{H},\textbf{s}\right)$ is equivalent to minimizing $\|\mathbf{y}-\mathbf{H}\mathbf{s}\|^2$, the JML detector is given by
\begin{align}
  \hat{\mathbf{s}} = arg \min_{\mathbf{s}\in{\Omega}}\|\mathbf{y}-\mathbf{H}\mathbf{s}\|^2.
\end{align}
Although the JML detector is an optimal detector, its complexity grows exponentially with the size of $\mathbf{s}$ and the modulation order. This motivates the design of suboptimal algorithms that reduce the detection complexity.
\subsubsection{Reduced Complexity List-Based Group-Wise Detector}
Referring to Fig.~\ref{fig:blockDiagram}, RC-LGSD consists of two entities: 1) branch list estimators (BLE) and 2) a global list optimizer (GLO). Although a similar approach to that of \cite{krause2011} is used for the processing within the BLEs and the GLO, the novelty of the proposed algorithm lies in its use of only $M$ BLEs instead of $N$. This lower number of BLEs reduces the complexity of the proposed algorithm with respect to \cite{krause2011}. This is made possible by the truncation process that was introduced in Section \ref{sec:truncation_noise}. The LGSD algorithm is briefly summarized below.

\begin{itemize}
  \item Let us denote the index set by $\Gamma=\{1, 2, ...,N\}=\left\{\gamma_1, \gamma_2,...,\gamma_U\right\}$ such that $\gamma_i \cap \gamma_j = \phi, \forall i\neq{j}$. We use the group index sets $\gamma_u$ to map different $s_i$, for $i=1, 2,\cdots, N$, to $\mathbf{s}_u$ and to map the equivalent channel matrix entries in the $m^{th}$ row, ${h}_{m,i}$, to $\mathbf{h}_u{(m)}$, for $m=1,\cdots,{M}$.
  \item The interference affecting the $u^{th}$ group in the $m^{th}$ BLE is cancelled by
\begin{align}\label{eq:IC}
  y_u^{(m)} = \mathbf{y}{(m)}-\sum_{v=1, v\neq{u}}^{U}\mathbf{h}_v\mathbf{s}_v,
\end{align}
$\mathbf{s}_v$ in (\ref{eq:IC}) is drawn from the best signal vector obtained in the previous iteration by the LGSD detector$^1$.\footnote{1 This is the first vector in the GLO output list $\mathbf{S}$. However, in the first iteration, $\mathbf{S}$ is populated from $\Omega$ at random.} (see Fig.~\ref{fig:blockDiagram}).

  \item The search is then performed by calculating the mean-squared error (MSE) for all $\mathbf{s}_u\in{\omega^{\mid\gamma_u\mid}}$ and $i=1,\cdots,{K^{\mid\gamma_u\mid}}$.
\begin{align}\label{eq:reduced_JML}
  e_u^{(i)}= \|y_u^{(m)}-\mathbf{h}_u(m)\mathbf{s}_u\|^2.
\end{align}
In this step, \textit{only} the $L$ $\mathbf{s}_u$ vectors with the lowest MSE are retained and mapped back to their branch candidate vector $\mathbf{s}_{br,m}^{(l)}$, $\ell=1,\cdots,L$, of size ($N\times{1})$, using $\gamma_u$.

  \item The branch candidate vector, $\mathbf{s}_{br,m}^{(l)}$, is sorted using
\begin{align}
  e^{(l)}(m) =  \|\mathbf{y}(m)-\mathbf{H}(m)\mathbf{s}_{br,m}^{(l)}\|^2,
\end{align}
to produce the branch list, $\textbf{S}_{br,m}=\{\mathbf{s}_{br,m}^{(l)}\}$.

  \item The $m^{th}$ BLE iterates over its own output $I_{BLE}$ times, by feeding back the produced $\mathbf{S}_{br,m}$ to its input. Afterwards, all $\mathbf{S}_{br,m}$ are forwarded to the GLO.
  \item In the GLO, all the branch lists are concatenated into a major list of size ($N\times{ML}$) that is again sorted by the MSE given by
 \begin{align}
  e =  \|\mathbf{y}-\mathbf{H}\mathbf{s}_{br,m}^{(l)}\|^2.
\end{align}
Only the first $L$ candidate vectors are preserved in a list, $\mathbf{S}_{in}$, while the rest are discarded.
  \item The rows of the list $\mathbf{S}_{in}$ are partitioned according to the mapping $\gamma_v$, for $v=1,\cdots,{V}$, resulting in $V$ lists with dimensions ($\mid\gamma_v\mid\times{L})$.
  \item The repeated vectors in these lists are dropped, resulting in $V$ lists with size ($\mid\gamma_v\mid\times{L_v})$, where $L_v\leq{L}$ is the number of the unique candidates in the $v^{th}$ list.
  \item The GLO uses these lists to find the highly probable candidate list $\mathbf{S}$.
  \item Subsequently, The GLO output list is fed back to its input for a further iteration.
  \item After $I_{GLO}$ GLO iterations, the final list is forwarded to the input of the BLE for a further global iteration. $I_{GLB}$ iterations are performed globally by the LGSD.
  \item A sorted list containing the highest probable candidates is generated at the output of the GLO and only the first vector is demodulated.
\end{itemize}
More details about the LGSD algorithm can be found in \cite{krause2011}.
\section{Complexity Analysis}\label{sec:complexity}
In this section, we provide a complexity analysis of the RC-LGSD and compare it to the complexity of the JML detector. We also show the complexity reduction compared to the LGSD algorithm due to the truncation process presented in Section \ref{sec:truncation_noise}. As a measure of complexity, we use a criterion similar to that of \cite{krause2011}, i.e., the number of real squaring operations required to obtain the Euclidean distance.

The computational complexity of the proposed RC-LGSD receiver can be calculated as
\begin{align}\label{eq:complexity_expression}
C(M) = 2I_{GLB}\left(MI_{BLE}\sum_{u=1}^{U}K^{\mid\gamma_u\mid}+I_{GLO}\sum_{v=1}^{V}{L_v\mid\gamma_v\mid}K^{\mid\gamma_v\mid}+R\right),
\end{align}
where $R$ is the number of the unique vectors in the major list at the input of the GLO. Note that, since $L_v$ is the number of unique vectors in the $v^{th}$ GLO sub-list, it varies from iteration to another, depending on GLO input list itself. Moreover, the first term in (\ref{eq:complexity_expression}) corresponds to the computational complexity of the BLEs, while the remaining two terms correspond to the complexity of GLO within the proposed detector.
\subsection{The Complexity of RC-LGSD Compared to JML}
In our investigations in Section \ref{sec:complexity_sim}, we provide a complexity/performance trade-off for the RC-LGSD receiver using different iteration sets. The complexity in this trade-off is provided as a percentage of the JML complexity required for the same system dimensions, i.e.,
\begin{align}
{C_{RC}=\frac{C(M)}{C_{JML}} \times{100\%}}
\end{align}
where $C_{JML} = 2MK^{N}$.
\subsection{The Complexity Reduction in RC-LGSD Relative to \cite{krause2011} and \cite{AbuShaban2013}}
In order to compare the complexity of the proposed RC-LGSD detector with the receivers in \cite{krause2011} and \cite{AbuShaban2013}, here, we compute the reduction in complexity for RC-LGSD as a percentage of the complexity of the schemes in \cite{krause2011} and \cite{AbuShaban2013}, that apply $N$ BLEs. This percentage is given by
\begin{align}\label{eq:comp}
C_{save} = \frac{C(N)-C(M)}{C(N)} \times 100\%.
\end{align}
Since $R<<NI_{BLE}\sum_{u=1}^{U}K^{\mid\gamma_u\mid}+I_{GLO}\sum_{v=1}^{V}{L_v\mid\gamma_v\mid}K^{\mid\gamma_v\mid}$, (\ref{eq:comp}) can be closely approximated by
\begin{align}\label{eq:save}
{C_{save} \cong \frac{(N-M)I_{BLE}\sum_{u=1}^{U}K^{\mid\gamma_u\mid}}{NI_{BLE}\sum_{u=1}^{U}K^{\mid\gamma_u\mid}+{I_{GLO}\sum_{v=1}^{V}{L_v\mid\gamma_v\mid}K^{\mid\gamma_v\mid}}} \times 100\%.}
\end{align}

Numerical results investigating the computational complexity of the proposed scheme and the approaches available in the literature are presented in Section \ref{sec:complexity_sim}.
\section{Simulation Results and Discussion}\label{sec:sim}
\subsection{Simulation environment}\label{sec:sim_env}
In this section, we present extensive simulations that determine the bit-error rate (BER) performance of the proposed receiver in practical settings and compare this performance to those of existing schemes available in the literature. The simulation setup for this paper can be summarized as follows:
\begin{enumerate}
\item We focus on modulation orders applied in satellite systems, namely, 8PSK and 16APSK shown in Fig.~\ref{fig:constellation}, which conform to DVB-S2 and DVB-Sx \cite{Morello2013}. The constellation radius ratio of 16APSK is selected to be 2.85, in accordance with \cite{dvbb-s2}.
\begin{figure}[!t]
\begin{center}
  \includegraphics[scale=0.4]{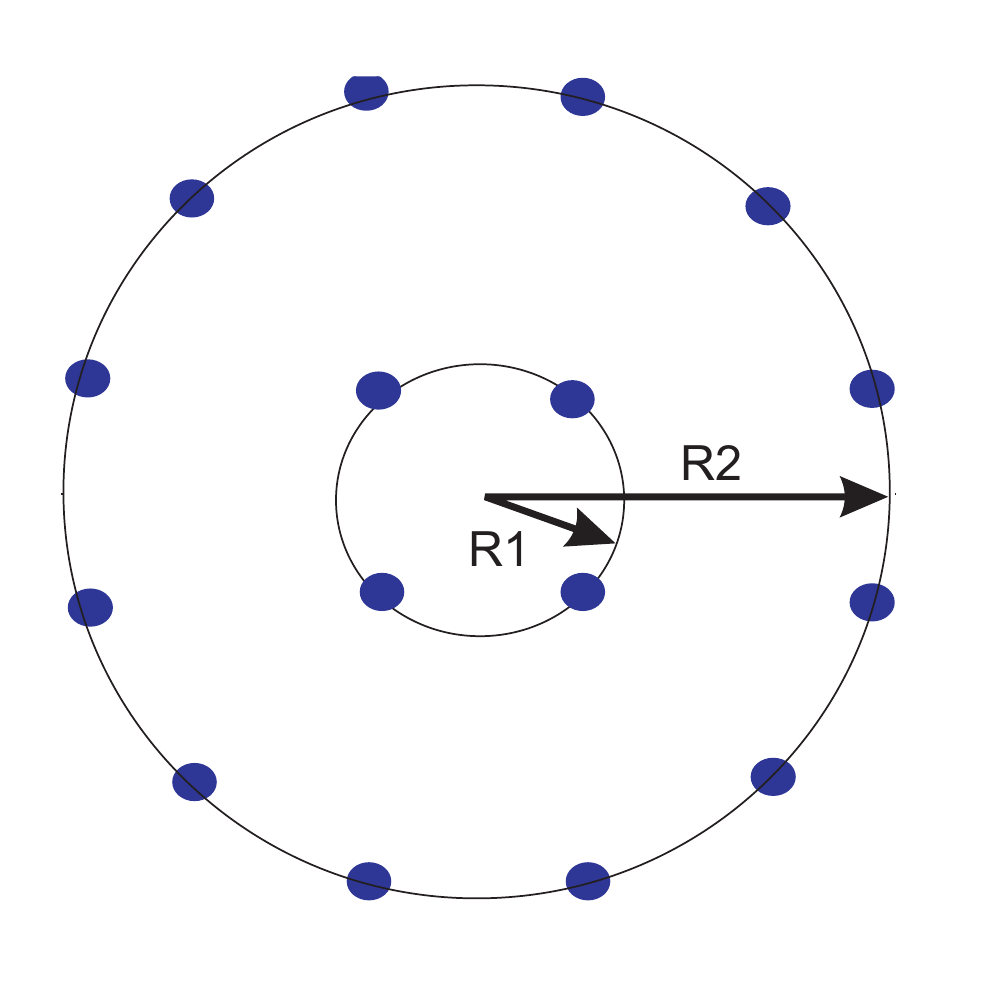}\\
  \vspace{-4mm}
  \caption{Constellation diagram 16APSK modulation. Radius ratio = $\frac{R_2}{R_1}$}\label{fig:constellation}
  \end{center}
\end{figure}
\item It is assumed that five GEO satellites, i.e., $N = 5$, that are stationed at orbital angles $\mathbf{\theta}=\left[\theta_n\right]=\left[0^\circ, 3^\circ, -2.8^\circ,5.7^\circ, -5.9^\circ\right]^T$ are in the view of the dish antenna. Here, $\theta_n$ is the orbital angle measured clockwise from the antenna axis as shown in Fig. ~\ref{fig:setup_figure}.

\item Recall that for a typical ASI scenario realization, adjacent satellites are assumed to be separated by $2.0^\circ-3.0^\circ$. The signals transmitted from the $5$ satellites are assumed to be co-channel signals in the Ku-band.

\item It is assumed that three LNBs ,i.e., $M = 3$, are used in conjunction with a $35$-cm dish antenna that is directed towards the desired satellite $s_1$.

\item The antenna radiation patterns are obtained using the satellite design software GRASP \cite{GRASP}, which accepts the dimensions and the frequency/wavelength as parameters. GRASP is widely used in the satellite research and professional teams due to its accurate and realistic models for parabolic antennas.

\item For the considered setup, we use a dimensionless version of the patterns in Fig.~\ref{fig:pattern} in conjunction with (\ref{eq:matrix-K}) to obtain
\begin{align}
\mathbf{K}_{nn}=\left(
\begin{tabular}{ccc}
  1.0 & 0.31 & 0.01 \\
  0.31 & 1.0 & 0.31 \\
  0.01 & 0.31 & 1.0 \\
\end{tabular}
\right).
\end{align}

\item Symbol SNR is defined in terms of the average received signal power to noise power, i.e., SNR = $\frac{\|A\|^2_F}{\sigma^2_nMN}$.

\item \item The main goal of the simulations is to detect the received signal from the desired satellite, since the dish is directed towards this satellite. It is also important to note that the received signal corresponding to the desired satellite is affected by the highest level of interference compared to the received signal from the remaining satellites. Thus, the BER results only for the desired satellite are provided.

\end{enumerate}
The notation LGSD($I_{GLB}/I_{BLE}/I_{GLO}$) and RC-LGSD($I_{GLB}/I_{BLE}/I_{GLO}$) are used to refer to the algorithm in \cite{krause2011} and the proposed receiver, respectively. Here, $I_{GLB}$ represents the number of global iteration performed in the LGSD receiver, $I_{BLE}$ denotes the number of iterations performed within the BLE, while $I_{GLO}$ denotes the number of iterations performed within the GLO. The choices of $I_{GLB}$, $I_{BLE}$, and $I_{GLO}$ significantly affect the behavior of the receiver as demonstrated in Section V-B. For group partitioning, we use two group index sets, $\gamma_1$ and $\gamma_2$ of sizes $3$ and $2$, respectively. This grouping is selected such that initially, the $3$ strongest signals, $s_1$, $s_3$ and $s_4$ in Fig.~\ref{fig:setup_figure} are mapped using $\gamma_1$, while the remaining signals are mapped using $\gamma_2$. In the subsequent iterations, the group allocation is made at random to diversify the detection process in each iteration. A similar mapping is also utilized in the BLE and the GLO. This grouping represents an acceptable trade-off between complexity and performance, since larger groups require searching over larger spaces, while by selecting smaller groups the advantages of joint processing diminish.

\subsection{Complexity reduction and performance trade-off}\label{sec:complexity_sim}
In this subsection, we investigate the effect of iteration numbers $I_{GLB}, I_{BLE}$, and $I_{GLO}$ on the receiver performance and the incurred complexity. Subsequently, we select a suitable trade-off between performance and complexity and compare the proposed receiver to that of \cite{krause2011} and \cite{AbuShaban2013} for 8PSK and 16APSK modulations.
\subsubsection{8PSK transmission}\label{sec:8psk_results}
\begin{figure}[!t]
\begin{center}
  \includegraphics[scale=0.6]{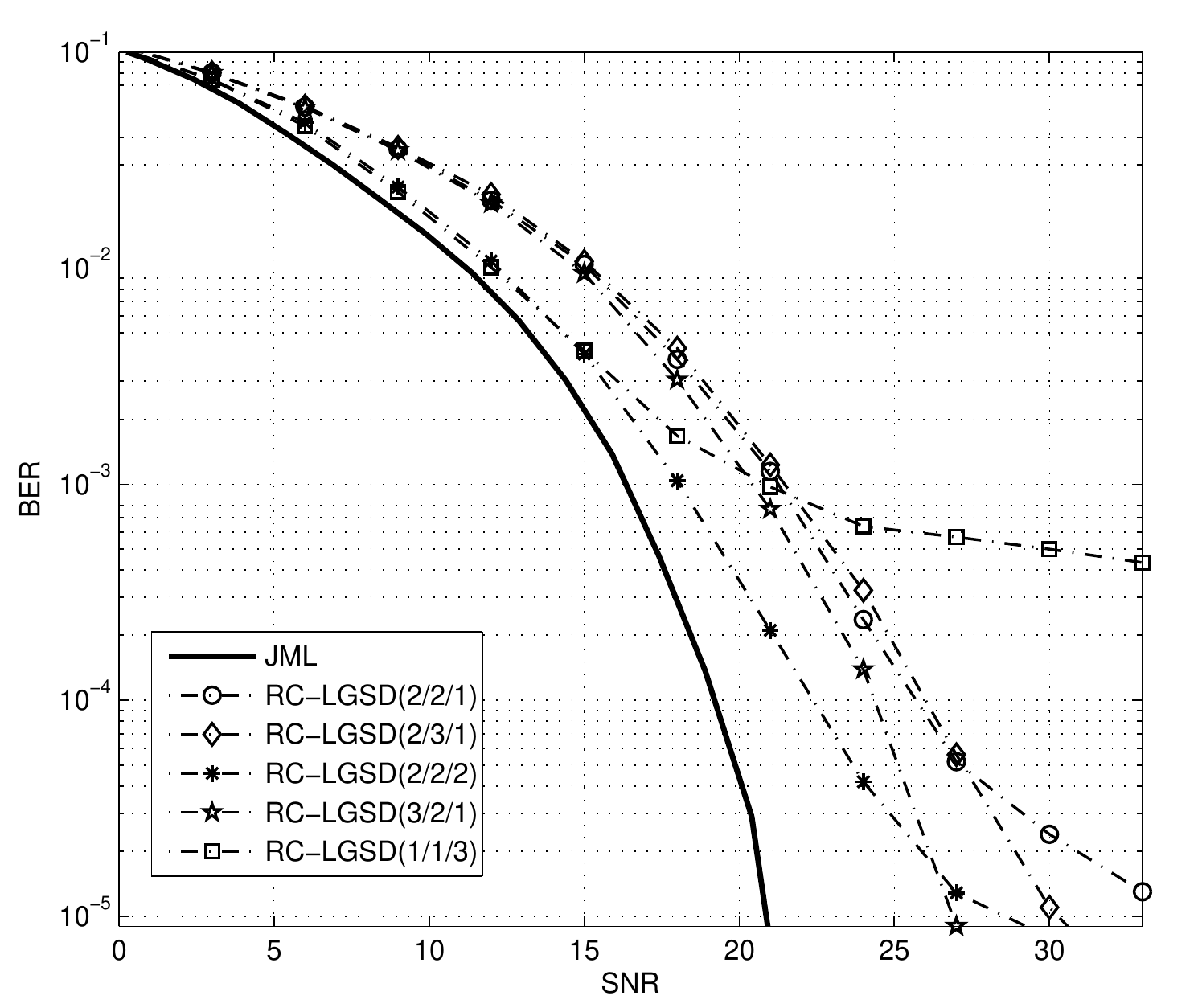}\\
  \vspace{-4mm}
  \caption{Performance of 8PSK using different iteration sets.}\label{fig:iterationEffect8}
  \end{center}
\end{figure}

\begin{table}\label{tabel:complexity}
 \begin{center}
  \caption{Complexity and required SNR of different iteration sets for 8PSK transmission at BER$=10^{-4}$}
  \renewcommand{\arraystretch}{1.2}
\begin{tabular}{|c|c|c|c|}
  \hline
$I_{GLB}/I_{BLE}/I_{GLO}$ & SNR(dB) & $C_{save}$ (\%)& $C_{RC}$ (\%)\\
  \hline
1/1/3& $\infty$  &2.7&42.4\\  \hline
2/1/1& 26  &7.1&30.6\\  \hline
2/1/2& 22  &3.9&57.7\\  \hline
2/1/3& 20  &2.7&84.7\\  \hline
2/2/1& 25.5  &12.1&34.1\\  \hline
2/2/2& 22.5  &7.1&61.2\\  \hline
2/2/3& 20.5  &5&88.2\\  \hline
2/3/1& 26  &15.7&37.6\\  \hline
2/3/2& 23.5  &9.8&64.7\\  \hline
2/3/3& 20.4  &7.1&91.8\\  \hline
3/1/1& 24.5  &7.1&45.9\\  \hline
3/1/2& 21  &3.9&86.5\\  \hline
3/2/1& 24.3  &12.1&51.2\\  \hline
3/3/1& 24.6  &15.7&56.4\\  \hline
JML & 19.5 & N/A & 100 \\
  \hline
\end{tabular}
\end{center}
\vspace{-0.8cm}
\end{table}
Fig.~\ref{fig:iterationEffect8} illustrates the performance of the receiver in terms of different number of iterations for 8PSK signals. In addition, Table I investigates the complexity of the proposed RC-LGSD approach with respect to LGSD, i.e., $C_{save}$, and the complexity of RC-LGSD with respect to the JML, i.e., $C_{RC}$, for different iteration values within the algorithm. Here, the SNR values are determined to reach the target BER value of $10^{-4}$. The analysis of Fig.~\ref{fig:iterationEffect8} and Table I leads to the following observations:
\begin{itemize}
  \item It can be inferred from Fig.~\ref{fig:iterationEffect8} that increasing the global iterations, $I_{GLB}$, enhances the performance of the receiver by eliminating the error floor at high SNR values. Moreover, for a larger $I_{GLB}$, the performance of the proposed RC-LGSD receiver comes closer to that of the JML. This can be illustrated by considering the two scenarios RC-LGSD(3/2/1) and RC-LGSD(2/2/1), where there is 2 dB saving in the low-to-medium SNR and the error floor eliminated at higher SNR. This comes with an additional cost of 17\% as indicated in Table I.

  \item Increasing the number of BLE iterations, $I_{BLE}$, barely enhances the performance of the receiver at low-to-medium SNRs. However, in this setup, the larger BLE iterations result in a lower error floor at high SNR values. This can be seen by comparing RC-LGSD(2/2/1) with RC-LGSD(2/3/1) in Fig.~\ref{fig:iterationEffect8}. This improvement in the overall performance of the system comes at the cost of overall receiver complexity, which is $3.5$\% higher for $I_{BLE}=3$ as shown in Table I.

  \item Increasing the number of GLO iterations, $I_{GLO}$, enhances the receiver performance by shifting the BER plot to the left, closer to that of the JML. This outcome is illustrated in Fig.~\ref{fig:iterationEffect8}, where by increasing $I_{GLO}$ from $1$ to $2$, for an additional complexity of $27$\%, RC-LGSD(2/2/2) results in a $3$ dB performance gain compared to RC-LGSD(2/2/1).

  \item Increasing the complexity of the proposed receiver does not necessarily reduce the required SNR to reach the BER of $10^{-4}$. This implies that the trade-off between performance and complexity for a given BER should be carefully selected by allocating different number of iterations to different blocks within the proposed receiver. This can be seen by noting that although the computational complexities of RC-LGSD(1/1/3) and RC-LGSD(2/3/1) are $42.4$\% and $37.6$\% of that of the JML detector, respectively, RC-LGSD(2/3/1) exhibits a superior performance as illustrated in Fig.~\ref{fig:iterationEffect8}.
\end{itemize}

It should be noted that, in some cases, the need for cheaper receivers surpasses the quality of service. Thus, Table I can be used to select a suitable trade-off between the affordable complexity and the associated performance.

\subsubsection{16APSK transmission}
\begin{figure}[!t]
\begin{center}
  \includegraphics[scale=0.675]{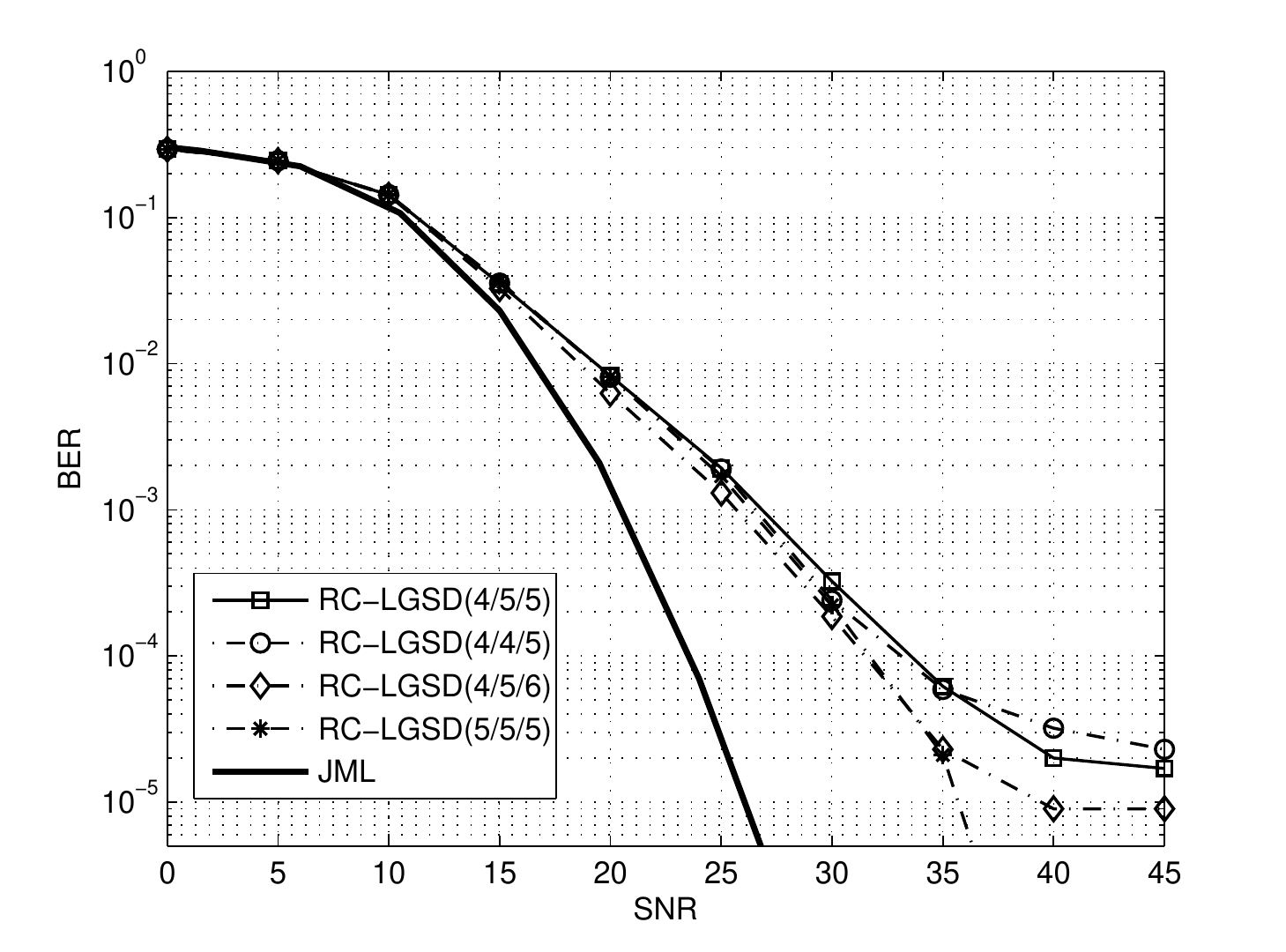}\\
  \vspace{-4mm}
  \caption{Performance of 16APSK using different iteration sets.}\label{fig:iterationEffect16}
  \end{center}
\end{figure}

\begin{table}
 \begin{center}
  \caption{Complexity and required SNR of different iteration sets for 16APSK transmission at BER$=10^{-4}$}
  \renewcommand{\arraystretch}{1.2}
\begin{tabular}{|c|c|c|c|}
  \hline
$I_{GLB}/I_{BLE}/I_{GLO}$ & SNR(dB) & $C_{save}$ (\%)& $C_{RC}$ (\%)\\
  \hline
4/4/4& 35.5  &7&56.6\\  \hline
4/4/5& 33.5  &5.8&69.1\\  \hline
4/4/6& 31  &5&81.6\\  \hline
4/5/4& 35.5  &8.4&58.2\\  \hline
4/5/5& 33  &7&70.7\\  \hline
4/5/6& 31.5  &6&83.3\\  \hline
4/6/4& 35.5  &9.6&59.8\\  \hline
4/6/5& 33  &8.1&72.4\\  \hline
4/6/6& 32  &7&84.9\\  \hline
5/4/4& 33.5  &7&70.7\\  \hline
5/4/5& 31.8  &5.8&86.4\\  \hline
5/5/4& 33.8  &8.4&72.8\\  \hline
5/5/5& 31.8  &7&88.4\\  \hline
5/6/4& 31.8  &9.6&74.8\\  \hline
6/4/4& 32  &7&84.9\\  \hline
6/5/4& 31  &8.4&87.3\\  \hline
JML & 24 & N/A & 100 \\
  \hline
\end{tabular}
\end{center}
\end{table}
The results of our investigations for determining the trade-off between performance and complexity for the proposed receiver while using 16APSK are presented in Fig.~\ref{fig:iterationEffect16} and Table II. Table II includes multiple scenarios with different number of iterations for each set within the proposed receiver, the associated complexity for each scenario, and the required SNR to achieve a BER of $10^{-4}$. Our findings can be summarized as follows:
\begin{itemize}
\item By increasing the global iterations, $I_{GLB}$, the error floor is significantly decreased at high SNRs, while for low-to-medium SNRs, the BER performance is closer to that of the JML detector. For example, when increasing $I_{GLB}$ from $4$ to $5$ and comparing RC-LGSD (5/5/5) with RC-LGSD(4/5/5), the overall performance of the system is enhanced by $1.2$ dB, while reaching a BER of $10^{-4}$. However, this added performance gain comes at the cost of an $18$\% increase in complexity. This is similar to the effect observed when increasing $I_{GLB}$ for 8PSK modulation in Section \ref{sec:8psk_results}.

\item On the other hand, removing one BLE iteration from the detection process does not greatly affects the performance of the proposed detector while greatly reducing its computational complexity. This can be observed when comparing the performance of RC-LGSD(4/5/5) with that of RC-LGSD(4/4/5) in Fig.~\ref{fig:iterationEffect16}, where the error floor at high SNR is slightly increased in the case of RC(4/4/5).

\item Our results indicate that, similar to the number of global iterations, increasing the number of GLO iterations can also significantly enhance the BER performance of the proposed receiver, while increasing its computational complexity. For example, RC-LGSD(4/5/6) is $13$\% more complex than RC-LGSD(4/5/5) but it is observed that it requires $1.5$ dB less than LGSD(4/5/5) to reach a BER of $10^{-4}$, while also reducing the error floor on the system performance.

\item Comparing the results in Tables I and II, it can be concluded that in contrast to 8PSK, 16APSK is more sensitive to interference since it is a two ring and denser modulation as shown in Fig.~\ref{fig:constellation}.
\end{itemize}

\subsubsection{Selecting Iteration Set}
The results in Fig.~\ref{fig:iterationEffect8} show that, the number of global iterations and GLO iterations, $I_{GLB}$ and $I_{GLO}$, respectively, mainly dictate the performance of the proposed RC-LGSD receiver. For example, when considering the RC-LGSD(2/2/1) scenario, the addition of a single iteration to the GLO stage enhances the BER performance of the system by $3$ dB while increasing the overall complexity of the receiver by $23.6$\%. On the other hand, the addition of a global iteration results in a $1.5$ dB performance gain while increasing the overall complexity of the system by $15.3$\%.

In conclusion, the global and GLO iterations have a major impact on the system performance, while larger values of BLE iterations, $I_{BLE}$, increase the overall computational complexity of the receiver without significantly enhancing the BER performance of the system. However, the number of BLE iterations dictate the BER error floor of the receiver. Thus, a moderate value of $I_{BLE}$ should be chosen. For instance, for 8PSK signals, our results indicate that RC-LGSD(2/1/2) is a reasonable compromise between performance and complexity since its complexity is $57.7\%$ of that of JML, while showing a $3$-dB poorer performance. \begin{figure}[!t]
\begin{center}
  \includegraphics[scale=0.6]{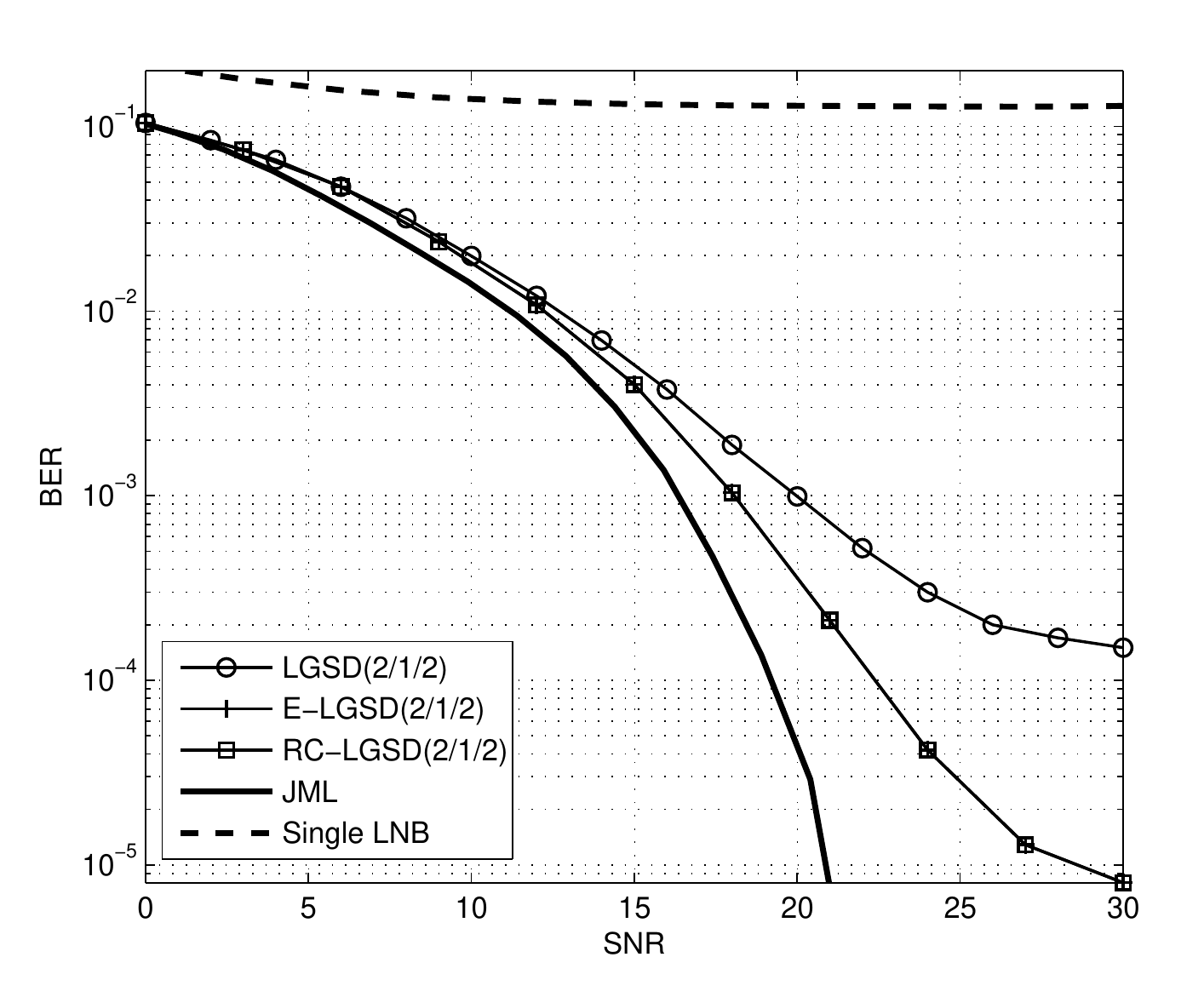}\\
  \vspace{-4mm}
  \caption{Performance of 8PSK using different detectors.}\label{fig:AbuShabanaKrause}
  \end{center}
\end{figure}Hence, it is used for comparison with existing schemes in the subsequent section and in Fig.~\ref{fig:AbuShabanaKrause}. With respect to 16APSK signals, our results show that RC-LGSD(4/4/5) presents a reasonable trade-off between complexity and performance, since its complexity is $69.1$\% of that of JML, while showing a $9.5$ dB poorer performance. Hence, RC-LGSD(4/4/5) is used in the next subsection for comparison with existing schemes.

\subsection{Performance Comparison}\label{sec:rc-lgsd_sim}
\begin{figure}[!t]
\begin{center}
  \includegraphics[scale=0.6]{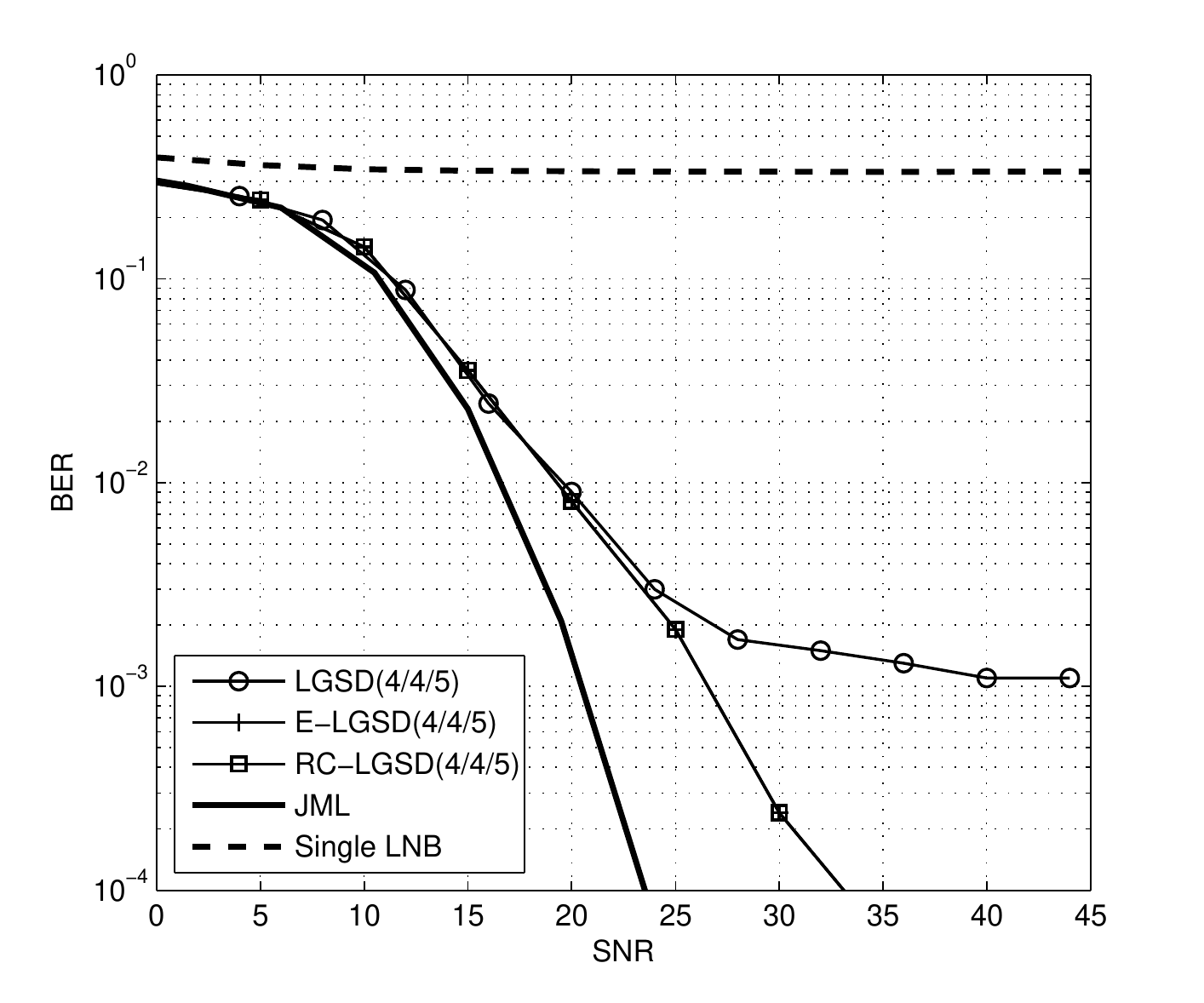}\\
  \vspace{-4mm}
  \caption{Performance of 16APSK using different detectors.}\label{fig:AbuShabanaKrause16}
  \end{center}
\end{figure}
A comparison of the performance of the proposed RC-LGSD receiver against that of the JML detector, the conventional LGSD \cite{krause2011}, and the enhanced LGSD (E-LGSD) \cite{AbuShaban2013} for 8PSK signals is presented in Fig.~\ref{fig:AbuShabanaKrause}. An iteration set of (2/1/2) is selected for this comparison as it constitutes a suitable performance and complexity trade-off. This figure shows that compared to conventional LGSD(2/1/2), the BER performance of RC-LGSD(2/1/2) is closer to the lower-bound presented by the JML detector by a considerable margin. Moreover, the error floor for RC-LGSD(2/1/2) is significantly lower when compared to conventional LGSD(2/1/2). This can be attributed to the application of the proposed beamformer that takes into account the structure of the interference for satellite systems. On the other hand, although RC-LGSD(2/1/2) and E-LGSD(2/1/2) \cite{AbuShaban2013} perform similarly, RC-LGSD is less complex with a $7.1$\% complexity saving as shown in Table I.

Considering Fig.~\ref{fig:AbuShabanaKrause16}, the same comment as above can be made regarding the performance of RC-LGSD and E-LGSD in 16APSK signals. The complexity saving for the iteration set (4/4/5) for E-LGSD is $5.8\%$ as compared to RC-LGSD. Moreover, when using the iteration set (4/4/5) for RC-LGSD and conventional LGSD, it can be observed that the error floor corresponding to RC-LGSD is significantly lower. This can be again attributed to the application of the proposed SINR-based beamforming approach in RC-LGSD compared to that of conventional LGSD in \cite{krause2011}.

Focusing on the application of single LNB in the presence of 3 satellites, the plots in Figs.~\ref{fig:AbuShabanaKrause} and~\ref{fig:AbuShabanaKrause16} show that the use of a single LNB by a receiver with a small-size dish antenna results in a poor BER performance. Consequently, adding two more LNBs provides spatial diversity at the receiver, which enhances the detection performance. This is in spite of the fact that the addition of more LNBs expands the field-of-view of the dish antenna and increases the interference at the receiver input (refer to Fig.~\ref{fig:pattern}).

\subsection{Pointing Error}
In this section we investigate the sensitivity of the proposed receiver and the JML detector to pointing errors. Pointing error may be caused by various factors such as dish misalignment, satellite drifting, and wind deflection. Let the pointing error vector be denoted by $\mathbf{\theta}'=[\theta_n']$. In accordance with \cite{Hughes1992}, $\theta_n'$ is modelled as a random variable drawn from a zero-mean normal distribution, $N(0,\sigma_e^2)$, where the error angle range is equal to $\theta_e=3\sigma_e$. Subsequently, the erroneous channel matrix can be written as a function of the satellite position and the error angle, which is denoted by $\mathbf{A}(\mathbf{\theta}+\mathbf{\theta'})$.
\begin{figure}[!t]
\begin{center}
  \includegraphics[scale=0.6]{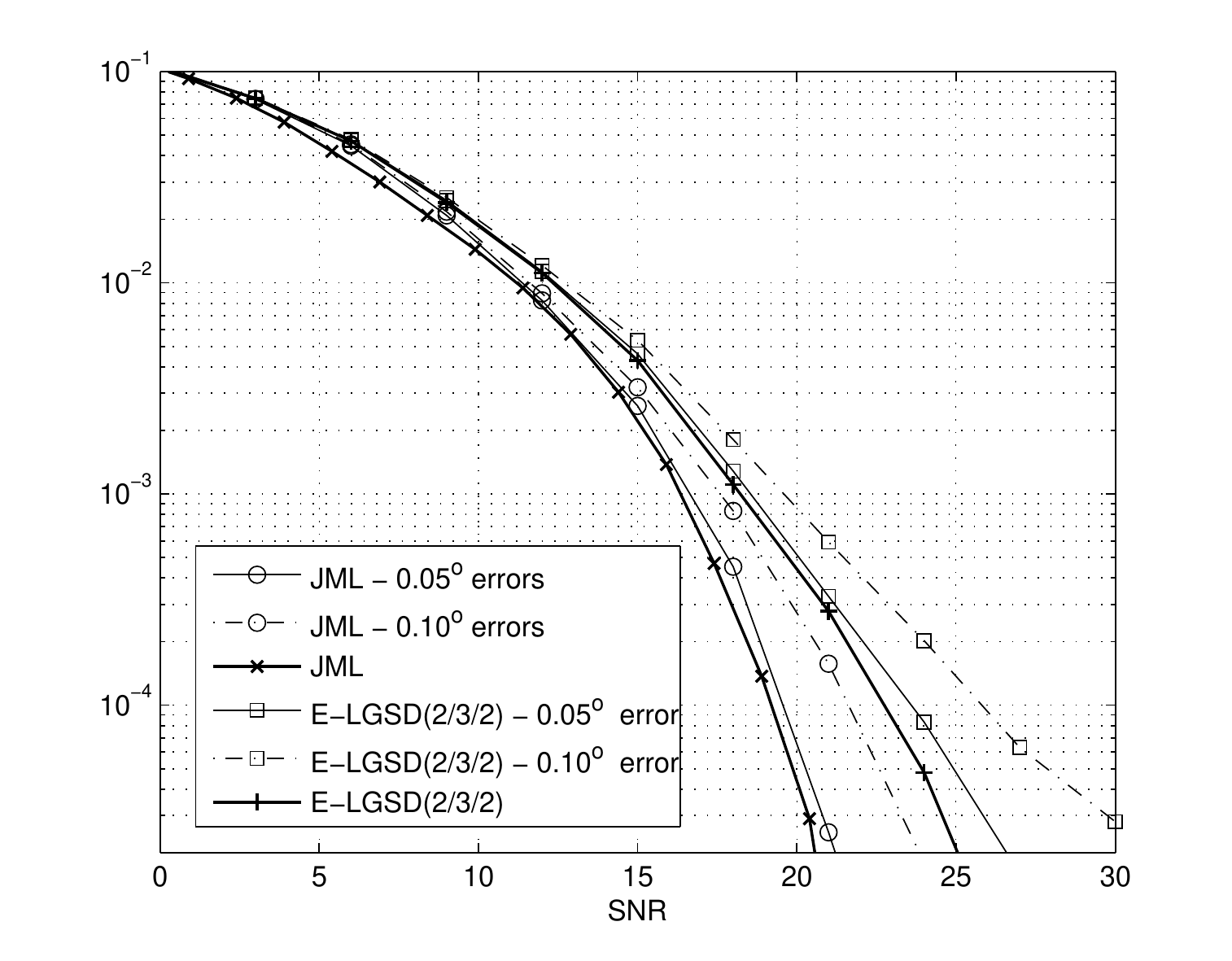}\\
  \vspace{-4mm}
  \caption{Pointing error effect on BER of the desired satellite, for error angle $0.05^\circ$ and $0.1^\circ$, and 8PSK modulation.}\label{fig:pointingError}
  \end{center}
\end{figure}

We assume that the procedure of \textit{beam bracket peaking} is used during the antenna steering \cite{Brooker2005}. This procedure is capable of pointing the dish antenna so that the largest error angle is $\theta_e=0.10^\circ$ and the median error angle is $\theta_e=0.05^\circ$. These two values result in different distributions for the pointing error, which are investigated separately here. The performance of the proposed receiver in the presence of pointing errors has been presented in Fig.~\ref{fig:pointingError}. It is inferred that the performance of the proposed detector deteriorates similarly to that of the JML detector in the presence of pointing errors. More specifically, it is observed that a pointing error of $0.10^\circ$ results in a $2.5-3.0$ dB performance loss when achieving a BER performance of $10^{-4}$. However, for a pointing error of $0.05^\circ$, a $0.5-1.0$ dB loss is observed at the same BER value. It should be stressed, however, that although the effect of pointing errors on the performance of satellite systems can be mitigated, the investigation of such mitigation is beyond the scope of this paper. Nevertheless, the effect of pointing errors can be alleviated via channel estimation using the pilot structure of the DVB-S2 signals, or using accurate apparatus during the mechanical setup.

\section{Conclusions}\label{sec:conc}
In this paper, we presented a reduced complexity LGSD receiver that modifies the linear preprocessor in the conventional LGSD receiver. To achieve this, an SINR-based beamformer is used instead of the MRC approach applied in prior work in the literature. In addition, a new filter for whitening the spatially correlated noise at the receiver input was derived. The enhanced receiver was applied to satellite broadcast systems in an overloaded setup. Furthermore, the computational complexity of the proposed receiver was further reduced via a channel truncation approach.

The simulation results were focused on the 8PSK and 16APSK modulations applied in satellite broadcast systems, to demonstrate the complexity and performance trade-off for the proposed receiver and to compare its performance with existing algorithms. It was indicated that the allocation of a higher level complexity to different receiver blocks should be done with caution, since a higher computational complexity does not necessarily result in better system performance. It was also shown that compared to the conventional LGSD approach the performance of the proposed receiver is closer to the optimal scenario presented by JML, while also reducing the computational complexity of the satellite receiver. In fact, for 8PSK and 16PSK we observe an SNR gain of $3$ and $13$ dB for a BER of $10^{-3}$, respectively. The proposed receiver exhibits a similar behavior to that of the JML in the presence of pointing error. It was observed that the largest pointing error, $0.10^\circ$, results in a performance loss of $2.5-3.0$ dB.


\end{document}